\documentclass[12pt]{article}
\usepackage{amssymb}
\usepackage{amsmath}
\usepackage{epsfig}
\newcommand{\be}{\begin{equation}}
\newcommand{\ee}{\end{equation}}
\newcommand{\bea}{\begin{eqnarray}}
\newcommand{\eea}{\end{eqnarray}}
\newcommand{\oha}{^1\!\!/\!_2}
\makeatother
\makeatletter
\renewcommand{\@biblabel}[1]{#1.}
\makeatother
\begin{document}
\title{Measurement of the energy of the 8.3-eV isomer of $^{229m}$Th with photoelectric effect.}
\author{F. F. Karpeshin}

\maketitle

\begin{center}
D.I.Mendeleyev Institute for Metrology VNIIM, Saint-Petersburg 190005, Russia \\ e-mail: fkarpeshin@gmail.com 
\end{center}

\begin{abstract} 
It is proposed to use the photoelectric effect in the inner shells of the $^{229}$Th atom to refine the energy of its 8.3-eV isomer. The calculation was performed using the Feinberg---Migdal shaking theory, which leads to the probability of isomer formation up to several units of $10^{-4}$ in the case of 
$K$-shell. As a result, two lines are predicted in the photoelectron spectrum,  their separation energy would provide the energy of the isomer. Other ways of using the method to study the properties of the isomer are also discussed: through shaking during the formation of radioactive beams in  storage rings, etc. Moreover, recording the effect in an experiment will make it possible to clarify its partial lifetime.
\end{abstract}

       At the time being, the most accurate atomic clocks are based on the hyperfine interaction of the electron shell of an atom with the nucleus. Record specimens achieve a relative error in reproducing a frequency (or time) unit within several times of $10^{-18}$ \cite{BSH}. This is not enough for fundamental purposes such as determining the drift of fundamental constants, detecting ultralight dark matter bosons \cite {BSH}, or for solving applied problems of navigation-coordinate support and others. The next generation of watches will use the nuclear isomeric line directly for stabilization. The nuclei, being located in the center of the electron shell, are less susceptible to external and intracrystalline fields. These lines are quite narrow and stable. The most likely candidate is currently the isomer $^{229}$Th, whose excited state $3/2^+[631]$ is located at a height of $\omega_n$= 8.338(24) eV (\cite{sandro}) above the ground $5/2^+[633]$.
Within the framework of this work, atoms of radioactive nuclides with $A$ = 229 were introduced into the crystal lattices of CaF$_2$, MgF$_2$, and others. As a result of beta decay, $^{229}$Th nuclei were formed, and even with a higher isomeric ratio as compared to alpha decay of $^{233}$U. The conversion decay channel was suppressed due to the presence of a band gap with a width exceeding the energy of the isomer. This made it possible to observe photons from nuclear decay and determine their energy.
The natural linewidth of the isomeric transition is only $10^{-5}$ Hz \cite{vityaf,arx}, which ssuits well the coherence time of laser devices available. 

       So far, however, the necessary condition for constructing a frequency standard is missing: the transition energy is unknown within the required error, which is determined by the natural width of the nuclear line. To determine it, the method of scanning the laser frequency in search for resonance photoexcitation of the isomer is applied. However, to have a chance for success, this approach must be built on the proper use of the resonant responce from the electron shell \cite{vityaf}. The most developed modern project is based on the use of a frequency comb to search for the resonant frequency in neutral atoms \cite{zhang}.
  The method uses the broadening of the isomeric line by 9 orders of magnitude due to internal conversion, otherwise the search for resonance would take an unrealistically long time for practical realization.

       The electron shell also provides the opportunity to use another method for determining the energy of the isomer, discussed below. First, however, let us briefly recall the basics of the interaction between the ground and isomeric levels of the nucleus through the electron shell.
For this purpose, let us dwell on the results of the work of \cite{Zyl}, in which the acceleration of the decay of the isomer in hydrogen-like ions was shown due to the mixing of the ground and isomeric levels of the nucleus in atomic states with a total angular momentum $F$ = 2. The coherent superposition of their amplitudes causes oscillation in the time of population of the ground and isomeric states of the nucleus around the average values. Let's look at these two questions in order.

In the ground state of such a hydrogen-like system, the spins of a single electron and the nucleus can be parallel or antiparallel. The corresponding total moment will be equal to $F$ = 3 or 2. The spin of an electron with an isomeric nucleus is added in a similar way, forming states with $F$ = 2 or 1. Thus, the state of an atom with $F$ = 2 can be of two types: 1) the nucleus is in the ground state, the spins are antiparallel, and 2) the nucleus is in the isomeric state, the spins are parallel. And since they interact through resonant conversion, then, according to the principles of quantum mechanics, these states form a superposition, which will be the real wave function. Denoting the mixing amplitudes of atomic states of the two types as $\alpha$ and $\beta$, respectively, the wave function of an atom in the ground and isomeric states can be written as
\bea
\Psi_g = \alpha|1\rangle +\beta|2\rangle  \nonumber \\
\Psi_\text{is} = \alpha |2\rangle - \beta |1\rangle\,,  \label{mx}
\eea
where $\alpha \approx$ 1, and $\beta$ can be determined through the matrix element of interaction and the energy difference between the two states in the first order of perturbation theory. If an atom is in the ground state, then the nucleus can nevertheless be in the isomeric state with probability $|\beta|^2$, and if the atom is in the isomeric state, then with the same probability its nucleus can be in the ground state. This is achieved due to the fact that the main and admixed amplitudes oscillate in time. Now the question of the transition of an atom from an isomeric state to the ground state, for example, a radiative one, can be viewed as follows: at some point in time, the nucleus is already in the ground state, so it is enough to spin-flip the electron, emitting excess energy. And since it is orders of magnitude easier for it to do this, as noted in the Introduction, the isomer's lifetime is reduced by hundreds of times. Another beautiful consequence of the Warsaw effect of mixing of nuclear levels for the gyromagnetic ratio of the $^{229}$Th nucleus in the ground state was drawn attention in the work \cite{shab}.

It seems no less tempting to use the oscillations in the ground state of the atom. For example, imagine that we suddenly remove the electron. Then the nucleus will evidently remain in that state in which it was at the moment of separation. This can be either a ground state or an isomeric state, and the ratio of probabilities is directly given by the ratio of the squared amplitudes of the population of both states. And due to conservation of energy, the energies of the photoelectrons will just differ by the energy separation of the nuclear doublet, up to the isomeric shift. We can thus determine the energy of the isomer by measuring the energy difference between the two photoelectron lines.

      The theory of effects arising in nuclear physics due to the sudden excitation and ionization of atoms was developed by Feinberg and independently by Migdal. There is a plenty of manifestation of such effects. They are observed in molecular, solid-state systems, playing a key role in experiments of the LUX \cite{LUX}, XENON1T \cite{XENON1T},
       DarkSide-50 \cite{DarkSide:2023},
       aimed at recording the interaction of dark matter particles with matter. The recoil of a nucleus due to interaction or a change in its charge, according to the Feinberg-Migdal theory, is accompanied by ionization and excitation of atoms, the detectors being oriented towards recording their  relaxation. In the case of double neutrinoless $e$-capture, taking into account the shake-up leads to its acceleration to an order of magnitude \cite{2K}. In double beta decay, a change in nuclear charge by two units is associated with excitation of the daughter atom in at least 75\% of decay cases, with an unexpectedly high mean excitation energy, ranging from 300 eV to 1 keV \cite{kriv}. To illustrate the idea in our case, consider the Feynman plot in Fig. 1 in the Furry's representation for the photoelectric effect on the valence $7s$ electron of a singly ionized $^{229}$Th atom. The nucleus transforms into an isomeric state as a result of the virtual exchange of a photon with a valence electron. The electron transfers  to an energy $\omega_n$ below the mass shell. It then absorbs a photon from the laser field and excapes the atom. 
\begin{figure}
\includegraphics[width=0.8\textwidth]{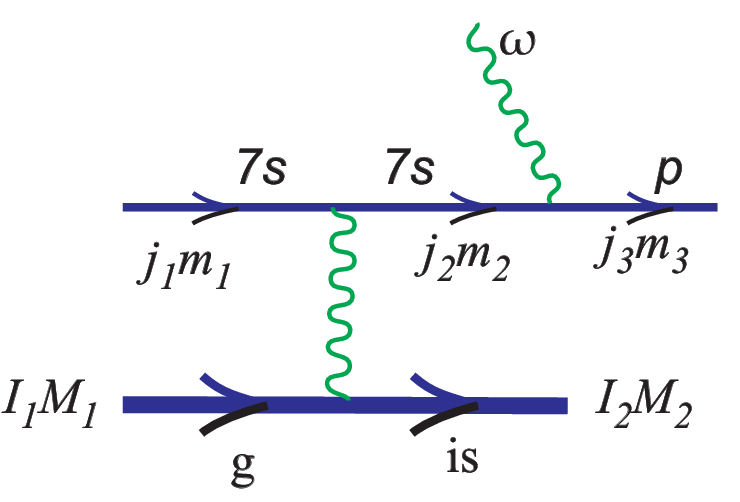}
\caption{\footnotesize Feynman graph of photoelectric effect on the valence $7s$ electron in the Th$^+$ ion, the nucleus left in the isomeric state.}
\label{f1}
\end{figure}
Baring in mind the Warsaw effect, the amplitude of the process can be expressed as follows:
\bea
F_1 = \sum_{j_2m_2M}\langle j_3m_3|H'_\gamma(\lambda,\mu)|j_2m_2\rangle
\frac1{-\omega_n}\langle I_2M_2j_2m_2||H'_c(LM)||I_1M_1j_1m_1\rangle
= \nonumber \\  \sum_{j_2m_2M}\frac{C(j_2m_2\lambda\mu|j_3m_3)}{\sqrt{2j_3+1}}
\frac{\langle j_3||H'_\gamma(\lambda)||j_2\rangle}{\omega_n}
\frac{C(j_2m_2  LM |j_1m_1)C(I_1M_1LM|I_2M_2)}{\sqrt{(2j_1+1)(2I_2+1)}} \nonumber \\
\times \langle I_2j_2||H'_c(L)||I_1j_1\rangle
\eea
Here $j_1, m_1, j_2, m_2, I_1, M_1, I_2, M_2$ are the angular momenta and their projections of the $7s$ electron and the nucleus, as shown in the graph in Fig. 1. The corresponding wave functions are normalized to one particle per unit volume, just like the spherical wave function of a photon with quantum numbers $\lambda \mu$. $j_3, m_3$ --- electron quantum numbers in the continuous spectrum. Its wave function, also in the form of a spherical wave, is normalized to the $\delta$-function in energy scale. To go to the probability of the process in Fig. 1, square of the amplitude must be averaged over the initial quantum numbers, summed over the final ones, and the result multiplied by $2\pi$. The result reads as follows:
\bea
W_1= \frac{2\pi}{(2\lambda+1)(2j_1+1)(2I_1+1)}\sum_{M_1M_2\mu m_1j_3m_3}|M_1|^2
= \nonumber \\
\sum_{j_3}\frac{2\pi|\langle I_2j_2||H'_c(L)||I_1j_1\rangle
\langle j_3||H'_\gamma(\lambda)||j_2\rangle|^2}
{(2\lambda+1)(2L+1)(2j_2+1)(2j_1+1)(2I_1+1)\omega_n^2} \,.
  \label{W1}
\eea
Since in the intermediate state we limited ourselves to the $7s$-level, then in (\ref{W1}) only the terms with $j_2=j_3=\oha$ are significant. The contribution to the amplitude of the process from a graph similar to Fig. 1, but with the reverse order of interaction of the $7s$ electron with the photon and the nucleus is much less (for example, compare the calculation of the graphs in Fig. 3, $a$ and $c$ from the Ref. \cite{npa}). To obtain the cross section of the photoelectric effect from the expression (\ref{W1}) for the probability, it must be divided by the flux of incident photons equal to $c$.
	\begin{figure}
\includegraphics[width=0.8\textwidth]{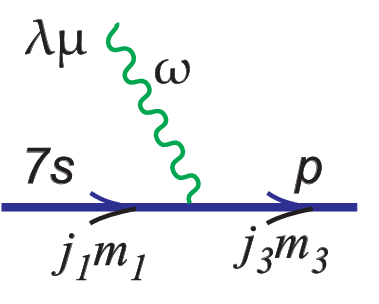}
\caption{\footnotesize Feynman graph of photoelectric effect in the valence $7s$ electron of the Th$^+$ ion.}
\label{f2}
\end{figure}

       The Feynman graph corresponding to the amplitude of the ordinary photoelectric effect is presented in Fig. 2. The amplitude reads as follows:
\be
M_2 = \langle j_3m_3|H'_\gamma(\lambda,\mu)|j_1m_1\rangle\,,
\ee  and the probability comes as follows:
\be
W_2 = \frac{2\pi}{(2\lambda+1)(2j_1+1)}\sum_{j_3}
|\langle j_3||H'_\gamma(\lambda)||j_1\rangle|^2  \,.  \label{W3}
\ee

The reduced conversion matrix element in Eq. (\ref{W1}) can be expressed in terms of the product of the discrete conversion coefficient and the radiative width of the nuclear transition (for example, \cite{kabzon,book})
\be
|\langle I_2j_2||H'_c(1)||I_1j_1\rangle|^2  = 3(2I_1+1)(2j_1+1)
\alpha_d(M1;j_1\to j_2) \Gamma_n^{(\gamma)}(I_1\to I_2)/2\pi \,,
\ee
As stated above, resonance conversion opens up the possibility of mixing the ground and isomeric nuclear states, despite the fact that they have different spins, while maintaining the total angular momentum of the $F$ atom. If in the diagram Fig. 1 consider hydrogen-like ions instead of single ones, then the total moments $F$ = 2 and 3 would be possible in the initial state, with the $F$ = 2 level  being the ground one. Similarly, an isomer with an electron could form the states with $F$ = 1 and 2. A virtual transition to the isomeric state is possible with $F$ = 2, while with the initial state with $F$ = 3 a process of the type in Fig. 1 would be impossible. In the case of single ions under consideration, due to the even larger electron shell moment $j = 3/2$, a whole series of combinations for the total moment $F$ both in the initial and final states is all the more possible. The selection rules for discrete conversion automatically take these combinations into account in the formulas for calculating discrete ICC (internal conversion coefficients) $\alpha_d(M1; 7s-7s)$. Now dividing (\ref{W1}) term by term with (\ref{W3}), we conclude that
\bea
W_1=\beta^2W_2\,, \nonumber \\
\beta^2=\frac{\alpha_d(M1;j_2\to j_1) \Gamma_n^{(\gamma)}(I_1\to I_2)/2\pi} {\omega_n^2}  \,. \label{gai}
\eea
The expression (\ref{gai}) is nothing more than a formula for the probability of mixing of nuclear states in (\ref{mx}) in the first order of perturbation theory, in full agreement with the previously expressed anticipation that the population of levels is proportional to the squares of the corresponding amplitudes in (\ref{mx}).

	\begin{figure}
\includegraphics[width=0.8\textwidth]{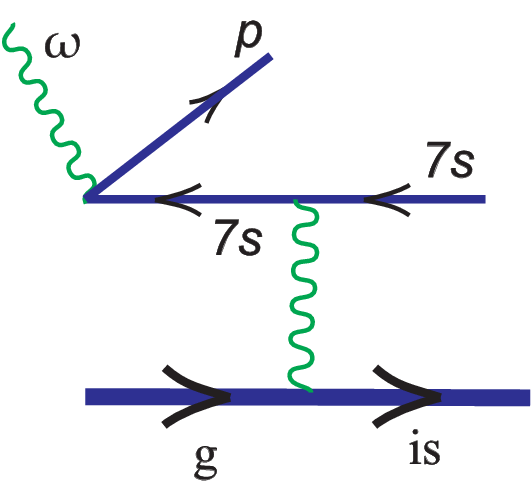}
\caption{\footnotesize Feynman graph of photoelectric effect on the valence $7s$ electron shell of neutral thorium atoms, the nucleus left in the isomeric state.}
\label{f3}
\end{figure}
The same effect of mixing of nuclear levels also occurs in neutral thorium atoms. The Feynman diagram of the excitation of the isomer during ionization of a $7s$ electron is shown in Fig. 3. An incident photon knocks out one of the $7s$ electrons, forming a hole in the shell with an excess of energy above the mass surface by $\omega_n$. When a hole interacts with a nucleus, this energy is transferred to the nucleus, transforming it into an isomeric state. The calculation of this diagram is quite similar to the one shown above and leads to the same result (\ref{gai}). The spectrum of electrons knocked out by a photon will also consist of two lines, separated exactly by the isomer energy $\omega_n$.

The last example suggests that the formation of a hole in the inner shells will also lead to the population of the isomer according to the Feynberg-Migdal mechanism. The difference appears in that the width of the hole state on the inner shells reaches a noticeable value: 88.2, 14.3 and 15.5 eV in the $K$-, $L_1$- and $M_1$-shells, respectively \cite{AD}. The presence of a width in the final state will manifest itself by the appearance of the same width in the photoelectron lines due to the conservation of energy. Mathematically this is expressed by adding the resonance 
Breit---Wigner factor in Eq. (\ref{W1}) for the electron spectrum:
\be
\frac{dW_1}{dE} =
\sum_{j_3}\frac{2\pi|\langle I_2j_2||H'_c(L)||I_1j_1\rangle
\langle j_3||H'_\gamma(\lambda)||j_2\rangle|^2}
{(2\lambda+1)(2L+1)(2j_2+1)(2j_1+1)(2I_1+1)\omega_n^2} \frac{\Gamma/2\pi}{(E-E_0)^2 + (\Gamma/2)^2}  \,,
  \label{W1c}
\ee    
where $\Gamma$ is the total width of the hole, $E$, $E_0$ is the kinetic energy of ejected electrons and its resonance value.

After integration over the electron line profile, provided that 
$E_0\gg \Gamma$, for the probability of isomer excitation we obtain the same expression (\ref{W1}), and
(\ref{gai}) --- for the branching ratio.

Discrete conversion coefficients have been calculated, for example, in refs. \cite{arx,npa}. For an isomer energy of 8.338 eV $\alpha_d(M1;7s\to 7s) = 3.90\times10^{10}$ eV, $\alpha_d(M1;1s\to 1s) = 5.14\times10^{18}$ eV, $ \alpha_d(M1;2s\to 2s) = 1.77\times10^{17}$ eV, and $\alpha_d(M1;3s\to 3s) = 7.84\times10^{15}$ eV for neutral atoms, respectively. By means of Eq. (\ref{gai}) one finds the probability of isomer creation
$\beta^2 = 2.62\times10^{-4}$, $0.90\times10^{-5}$, $4.00\times10^{-7}$ and $1.99\times10^{-12}$ for the $1s$- , $2s$-, $3s$- and $7s$-states, respectively. The results are also presented in the Table.
\begin{table}
\caption{\footnotesize Discrete ICC $\alpha_d(M1;ns\to ns)$, isomer creation probabilities $\beta^2$ and total hole widths $\Gamma$.}
\begin{center}
\begin{tabular}{||c||c|c|c|c||}
\hline \hline
& $1s$  &  $2s$  & $3s$  &  7s  \\
\hline
$\alpha_d(M1;ns\to ns)$, eV  &  $2.57\times10^{18}$  &  $0.89\times10^{17}$  &  $3.92\times10^{15}$  &  $0.98\times10^{10}$  \\
$\beta^2$  &  $2.62\times10^{-4}$  &  $0.90\times10^{-5}$  &  $4.00\times10^{-7}$  &  $1.99\times10^{-12}$  \\
$\Gamma$, eV \cite{AD}  & 88.2 &  14.3  &  15.5  & --- \\
\hline \hline
\end{tabular}
\end{center}
\end{table}

Our results for the $1s$ electrons are consistent with those obtained earlier in \cite{Zyl,npa}. If one puts the energy of the isomer in Eq. (\ref{gai}) $\omega_n$ = 3.5 eV and take into account that $\alpha_d(M1;ns\to ns)\sim \omega_n^{-3}$, then he arrives at the probability of  $ \beta^2$ = 0.02, which is the same as that used in Ref. \cite{Zyl}. A similar agreement is with Ref. \cite{npa}. For the 
$2s$-electrons, the discrete ICC and, accordingly, the probability of excitation are an order of magnitude lower. This is to be expected since, to a first approximation, the coefficients are proportional to the electron density at the origin. The decrease is also enhanced due to electron screening. For the upper shells the effect accordingly. Thus, the effect of isomer excitation during atomic shaking is maximal for the $K$-electrons, and with increasing shell number it decreases. On the other hand, the width of the electron line profile is equal to the width of the hole. Therefore, it is easier to experimentally resolve two energy lines in the case of hole formation in upper shells. Consequently, when choosing an experimental technique, one must proceed from specific conditions, goals and objectives.

We summarize that the avalanche-like increasing waive of  the number of experimental papers where the energy of the isomer is consistently clarified (for example, \cite{sandro,larsrv} and Refs. cited there), as well as new projects \cite{cB,cBH} suggests that the main goal of  the contemporary research into the problem of $^{229m}$Th --- determining its energy --- achieves the decisive phase.
Against this background, the method proposed above for measuring the isomer energy can significantly reduce the uncertainty. Regardless of which shell the photoelectric effect is observed in, the photoelectron spectrum predicts coupled lines separated by the isomer energy. Of course, the excitation of the isomer due to the sudden formation of a hole in the shell is of interest in itself. It is appropriate to note that the probability of isomer formation is proportional to the partial width of its radiative decay. Therefore, studying the effect will help to refine its meaning.

       In addition to the photoelectric effect, sudden ionization can be produced by electron impact and other methods. The Ref. \cite{npa} considers ionization during the passage of heavy ion beams through light (beryllium) targets. This technique was just tested in the storage rings at GSI Darmstadt, ISF Lanzhou. It is a common tool for generating radioactive heavy ion beams. Moreover, according to the above scheme, the photoelectric effect can be used in various combinations with other methods, for example, in experiments with a crystal lattice of the type described in Ref. \cite{sandro}. 

\bigskip

The author would like to express his gratitude to J. Zylicz and M. Pf\"utzner for initiating discussions.

\end{document}